\documentclass[a4paper,12pt]{article}

\usepackage[T2A]{fontenc}
\usepackage[english]{babel}
\usepackage{amsmath}
\usepackage{cite}
\usepackage{tikz}
\usepackage[unicode,linktocpage=true,plainpages=false,pdfpagelabels=false]{hyperref}

\newcommand{\ls}{\left(}
\newcommand{\rs}{\right)}
\newcommand{\ff}{\varphi}
\newcommand{\al}{\alpha}
\newcommand{\be}{\beta}

\newcommand{\disn}[2]{$$\displaylines{\refstepcounter{equation}%
            \label{#1}\hskip 1em minus 1em #2\hfilneg}$$}
\newcommand{\nom}{\hfil\hskip 1em minus 1em (\theequation)}
\newcommand{\no}{\hfil \hskip 1em minus 1em\phantom{(\theequation)}%
            \hfilneg\cr\hfilneg\hskip 1em minus 1em\hfil}
\newcommand{\ns}{\hfill\cr\hfill}

\begin{document}

\title{Analytical analysis of the origin of core-cusp\\ matter density distributions in galaxies}

\author{
A.~D.~Kapustin\thanks{E-mail: sashakapusta96@gmail.com},
S.~A.~Paston\thanks{E-mail: pastonsergey@gmail.com}\\
{\it Saint Petersburg State University, Saint Petersburg, Russia}
}
\date{\vskip 15mm}
\maketitle

\begin{abstract}
We propose an analytical method to describe a matter density profile near a galaxy center.  The description is based on the study of the distribution function of particles over possible trajectories. We establish a relation between the central slope of density profile and the near-origin behavior of the angular momentum distribution function.
We consider both a spherically symmetric (on average) matter distribution as well as deviations from it.
If the density profile forms in a  background of spherical gravitation potential then a core-type distribution arises.
A regular matter may behave in such way if the background potential was formed by the dark matter.
In the presence of deviation from spherical symmetry the formation of cusp-type distribution is possible.
Moreover, a reduction of spherical symmetry to the axial one leads to a less steep cusp profile. The complete symmetry breaking (which corresponds, in particular, to the common setup of numerical simulations), leads to a steeper cusp profile.
\end{abstract}

\newpage

\section{Introduction}
A modern theory describing large-scale universe evolution is the $\Lambda$CDM model. This model provides a satisfactory fit to the majority of observations related to the formation and evolution of cosmic structures on large scales \cite{DEL_POPOLO_2014, Spergel_2003, Komatsu_2011}. However, several
problems still remain unresolved, in particular, the \emph{core-cusp} problem, which is related to the  behavior of the density profile near the galaxy center.
In the observed dark matter density distribution near galaxy centers density profiles preserve smoothness in the center and form a \emph{core}. In the numerical simulations of the cosmic structure formation, the  initial matter distribution evolves into a singular density profile called  \emph{cusp}. The discrepancy between these facts is called the core-cusp problem.

The classic result of numerical simulations is the universal density profile obtained in works \cite{Navarro_1996, Navarro_1997} and frequently referred to as the NFW profile. In this profile the density of matter is singular in the center and proportional to $r^{\alpha}$, where $\alpha = -1$. This profile is stably reproduced in simulations of cosmic structures formation with non-interacting dark matter.
It is possible to suggest the mechanisms \cite{10.1093/mnras/283.3.L72} relying on star formation processes and gas dynamics, which can efficiently smooth the NFW distribution. However, these mechanisms are unable to resolve the core-cusp problem in the case of galaxies, where dark matter is a strongly dominant component.

We can pick a set of dwarf satellite galaxies of the Milky Way and Andromeda galaxy as candidates for a role of observed galaxies with a dominant dark component, where both gas-rich galaxies with ongoing baryonic processes and gas-poor galaxies are present. Initially, the core-cusp problem was found in the study of rotation curves of gas-rich dwarf galaxies \cite{Moore1994, 1994ApJ...427L...1F,Burkert_1995}, where the density profile differs from $1/r$ and corresponds to the core with constant density in the center, which means that density profile is proportional to $r^\alpha$, with $\alpha = 0$. More modern observations show less unambiguous results regarding the central structure of density profiles. One can find a thorough description of the core-cusp problem both from the side of observations and numerical simulations and a comprehensive list of references in a review \cite{1606.07790} or in a recently published review \cite{galaxies10010005}.

One possible way to look for a solution to the core-cusp problem is to introduce a self-interaction for dark matter (see \cite{1705.02358} for examples) because self-interaction may prevent excessive accumulation of dark matter in the center of a galaxy. It is worth noting that no one has succeeded in direct detection of dark matter \cite{1509.08767,1604.00014} in a framework of various hypotheses regarding the interaction of dark and regular matter (see, for example, \cite{1703.07364,astro-ph/0003365,1705.02358}). This leads to a possibility that dark-matter-related effects may be explained by a modification of GR, which would contain \emph{extra solutions}. Such solutions can be treated as solutions of GR with an additional contribution from a fictitious matter in the r.h.s. of Einstein equations.  In such a framework, the dark matter has a purely gravitational origin instead of being a regular matter, so it explains failures of its direct detection.

We can consider a popular model of mimetic gravity \cite{mukhanov,Golovnev201439} (see also review \cite{mimetic-review17}) as an example of a theory containing extra solutions along with all GR ones. Alternatively, we can use another theory called \emph{embedding gravity} (or \emph{embedding theory}) \cite{regge,deser}, which is similar to the mimetic gravity since both theories appear as a result of differential field transformations in GR (see details in \cite{statja60}). The dark matter arising in this approach in the non-relativistic limit behaves as a dust-like matter with some self-interaction \cite{statja51,statja68}. Therefore in this approach we can try to explain dark-matter-related effects not only on the cosmological scales \cite{davids01,statja26}, but also on the scales of galaxies.

Regardless of whether dark matter is the real matter with self-interaction, or a gravitational effect, it is necessary to evaluate how the presence of self-interaction affects the emerging profile of the distribution of matter in the galaxy -- would \emph{core} or \emph{cusp} appear.
In order to avoid resource-consuming simulations, we aimed to obtain analytic criteria of when \emph{core} and when \emph{cusp} distributions arise. The purpose of the current paper is to consider the case without any self-interaction. Particular cases with self-interaction of the dark matter could be considered in the future.

The applicability of purely analytical approaches to formation of cosmic structures is strongly limited, so spherically symmetrical problems are often considered. One of the simplest and most popular models is an isothermal sphere, which is described, for example, in \cite{Doroshkevich_2012,10.1046/j.1365-8711.1999.02609.x}. Pseudo-isothermal sphere with density profile $\rho \sim \left(1+ r^2/r_0^2 \right)^{-1}$ gives a flat density profile which appears to be in good agreement with observations.
An analytical approach to the dynamics of galaxy formation processes was proposed, for example, in \cite{https://doi.org/10.48550/arxiv.astro-ph/0006184,10.1046/j.1365-8711.1999.02609.x, https://doi.org/10.48550/arxiv.astro-ph/0409173, El_Zant_2008, El_Zant_2013}. In these papers the formation of the cosmic structure from the initial fluctuation and the stability of this process are discussed.
Another class of analytical problems, with a setup similar to ours, are works devoted to the study of the relation between a distribution function of particles in velocity space with a density profile for a static cosmic structure (see, for example, \cite{1985AJ.....90.1027M,refId0}).

In the present work we suggest a simple analytical approach for obtaining a central density slope of  a dust-like matter in a galaxy by considering the distribution function of particles over all possible trajectories.
Primarily, we are focused on a static and spherically symmetric (on average) distribution in the assumption that the movement of a single particle can violate spherical symmetry. In a section \ref{razd1} we obtain a relation between an arising profile type (\emph{core} or \emph{cusp}) and a near-origin behavior of the distribution function over a module of angular momentum.
In a section \ref{razd21} we show that for an exactly spherically symmetric gravitational potential this asymptotic corresponds to the \emph{core} profile. Taking this into account, the \emph{cusp} profile can arise only as a result of deviation from exact spherical symmetry. In a section \ref{razd22} we discuss the time evolution of the near-origin behavior of a distribution function over the angular momentum modulus, which leads to a transition of a density profile to the \emph{cusp}-type.
In a section~\ref{razddop} we discuss how the deviations from spherical symmetry affect the obtained results.

\section{Relation between the density profile and the distribution function of particles}\label{razd1}
We are interested in a matter distribution with exact spherical symmetry. Let us consider a static and spherically symmetric (on average) distribution of matter. We assume that the matter consists of massive particles which interact with each other only by means of gravity and move along finite trajectories. Let us also assume that particles move with non-relativistic speed and their density is sufficiently small so that Newtonian gravity can be applied.
In that case, a matter density is given by a time-independent spherically symmetric function $\rho$, which is related to the static spherically symmetrical gravitational potential $\varphi$ by the Poisson equation
\begin{equation}
\label{poisson}
	\partial_k \partial_k \varphi(x_i) = 4 \pi G \rho(x_i),
\end{equation}
where $G$ is the Newtonian gravitational constant; hereinafter $i,k,\ldots=1,2,3$.

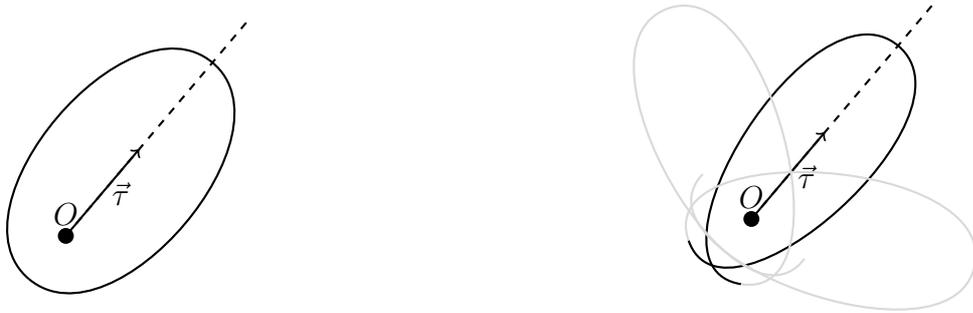
\begin{figure}[h!]
\begin{minipage}{0.49\linewidth}
	\centering
	\begin{tikzpicture}[scale = 1.5]
		\begin{scope}[rotate = -40]
			\draw[thick] (0, -0.5) to[out = 0, in = 0] (0, 2);
			\draw[thick] (0, 2) to[out = 180, in = 180] (0, -0.5);
			\draw[thick, dashed] (0, 0) -- (90:2.5);
			\draw[thick, ->] (0, 0) -- (90:1)node[pos = 0.75, below]{$\vec{\tau}$};
		\end{scope}
		\fill (0, 0)node[above]{$O$} circle(0.07);		
	\end{tikzpicture}
\end{minipage}
\hfill
\begin{minipage}{0.49\linewidth}
	\centering
	\begin{tikzpicture}[scale = 1.5]
		\begin{scope}[rotate = -40]
			\draw[thick] (-0.3, -0.5) to[out = -31, in = 0] (0, 2);
			\draw[thick] (0, 2) to[out = 180, in = 211] (0.3, -0.5);	
			\draw[thick, dashed] (0, 0) -- (90:2.5);
			\draw[thick, ->] (0, 0) -- (90:1)node[pos = 0.75, below]{$\vec{\tau}$};
		\end{scope}
		\fill (0, 0)node[above]{$O$} circle(0.07);
		\begin{scope}[rotate = 22, gray!30]
			\draw[thick] (-0.3, -0.5) to[out = -31, in = 0] (0, 2);
			\draw[thick] (0, 2) to[out = 180, in = 211] (0.3, -0.5);	
		\end{scope}
		\begin{scope}[rotate = -102, gray!30]
			\draw[thick] (-0.3, -0.5) to[out = -31, in = 0] (0, 2);
			\draw[thick] (0, 2) to[out = 180, in = 211] (0.3, -0.5);	
		\end{scope}
	\end{tikzpicture}
\end{minipage}
\caption{A definition of the orbit and its characteristic direction for a closed and open orbit.}
\label{pic1}
\end{figure}

When moving in a spherically symmetric potential, each particle has a conserved total energy $E$ and angular momentum $L_k$. In this case, each particle moves along a planar orbit whose plane contains the center of symmetry. Since we are considering a finite motion only, the change in the radial coordinate will be periodic, but the orbit may or may not be closed. In the case of an open orbit, by orbit we will, for definiteness, mean only the part corresponding to one period of radial motion (see fig.~\ref{pic1}).

Then each orbit can be uniquely defined with normalized (divided by mass of the particle $m$) energy $\varepsilon = {E}/{m}$, normalized angular momentum $\ell_k = {L_k}/{m}$ and a direction $\tau_k$, defining the orientation of the orbit in a plane orthogonal to the vector of angular momentum $L_k$.
Hence, the vector $\tau_k$ must satisfy conditions
\begin{equation}\label{sp1}
\tau_k \ell_k=0,\qquad \tau_k\tau_k=1
\end{equation}
and because of that it has only one independent component.
To describe a particle motion in full we need to introduce another scalar parameter $\gamma$ which defines a phase of a particle's periodic movement taken at the initial moment of time.
As a result, the motion of a single particle is given by the function
$\hat{x}_m \left( t, \varepsilon, \ell_k, \tau_l, \gamma \right)$, where $t$ is time.

Now we can introduce the distribution function of particles $f$ (without the loss of generality we can assume that all particles have the same mass $m$), which depends on all mentioned parameters. This function is defined by the number of particles within a small range of parameters according to the formula
\begin{equation}
	d N = f \left( \varepsilon, \ell_k, \tau_l, \gamma \right) \, d \varepsilon \,d^3 \ell \,d \tau \,d \gamma,
\end{equation}
where $d\tau$ is considered one-dimensional since the vector $\tau_l$ must satisfy conditions \eqref{sp1}.
To be more accurate, $d\tau$ should be defined as a product $d^3\tau\,\delta(\tau_k \tau_k-1)\delta(\tau_k \ell_k/\ell)$, where $\ell=\sqrt{\ell_k\ell_k}$.
In this section, we will assume that the process of galaxy formation has already been completed so the distribution function $f$ has become time-independent.

Let us derive the expression for the density of matter at a given point using the distribution function $f$. Considering the fact that contribution to the density from a single particle can be written using $\delta$-function, we can write the expression for the density in the form
\begin{equation}
\label{rho}
	\rho(x_m) = m \int d \varepsilon \,d^3 \ell \,d \tau \,d \gamma \ f \left(  \varepsilon, \ell_k, \tau_l, \gamma  \right) \delta \left( x_m -  \hat{x}_m \left( t, \varepsilon, \ell_k, \tau_l, \gamma \right) \right).
\end{equation}
The supposed spherical symmetry and time independence of the matter density $\rho$ must be ensured by some properties of the distribution function $f$, however, one can avoid their explicit specification.
Using the fact that l.h.s of the equation \eqref{rho} is independent of time and of a particular direction of the vector $x_m$, we can integrate both parts of the equation over time from $0$ to $T$ and over a sphere with radius $r=\sqrt{x_k x_k}$ and after that divide the result by the sphere surface area and by the length of the time interval $T$.
As a result, we obtain
\begin{equation}
\label{rho2}
	\rho(r) = \frac{m}{4\pi r^2 T} \int d \varepsilon \,d^3 \ell \,d \tau \,d \gamma \ f \left(  \varepsilon, \ell_k, \tau_l, \gamma  \right) \int \limits_{S_r} d^2 x \int \limits_0^T dt \, \delta \left( x_m -  \hat{x}_m \left( t, \varepsilon, \ell_k, \tau_l, \gamma \right) \right).
\end{equation}
The delta-function in this expression can be removed if we switch from integration over $t$ to integration over $r$ (obtaining an additional multiplier $1/|v_r|$, where $v_r$ is a radial component of velocity) and combine this integration with the integration over sphere $S_r$. This results in
\begin{equation}
\label{rho3}
	\rho(r) = \frac{m}{4\pi r^2 T} \int d \varepsilon \,d^3 \ell \,d \tau \,d \gamma \ f \left(  \varepsilon, \ell_k, \tau_l, \gamma  \right)
\frac{n}{|v_r|},
\end{equation}
where $n$ is a number of $\delta$-peaks which lie within the interval of integration over time. This number depends only on $T$ and $\varepsilon$, $\ell$ for any given $r$.

To proceed, let us return to the movement of a single particle. While moving in the spherically symmetric field, it possesses a conserved energy
\begin{equation}
	E  = \frac{m}{2} \left( v_r^2 + v_\tau^2 \right) + m \varphi(r)
\end{equation}
(here $v_\tau$ is a tangential component of velocity)
 and angular momentum $L_k$, whose magnitude has the form
\begin{equation}
	L =m r v_\tau.
\end{equation}
It allows us to express the module of the radial velocity through integrals of motion $\ell=L/m$ and $\varepsilon=E/m$:
\begin{equation}
\label{v}
	|v_r| = \sqrt{2\varepsilon - 2\varphi(r) - \frac{\ell^2}{r^2}}.
\end{equation}
The obtained expression can be used in \eqref{rho3}.

The number of $\delta$-function carriers $n(\varepsilon, \ell, r)$ in \eqref{rho3} can be estimated by introducing the period of radial motion
$\hat{T} \left( \varepsilon, \ell_k \right)$ which depends only on normalized energy and angular momentum:
\begin{equation}\label{sp3}
	n\left( \varepsilon, \ell, r \right) \approx \frac{2 T}{\hat{T} \left( \varepsilon, \ell_k \right)} \Theta \left( 2\varepsilon - 2\varphi(r) - \frac{\ell^2}{r^2}\right),
\end{equation}
where $\Theta(x)$ is the Heaviside step function. Moreover, the accuracy of the approximate equality \eqref{sp3} increases with an increase in the arbitrary parameter $T$.
As a result, using \eqref{v} and \eqref{sp3} in \eqref{rho3}, we obtain in the limit of large $T$
\begin{equation}\label{rho4}
\rho(r) = \frac{m}{2\pi r^2} \int d \varepsilon \, d^3\ell\,
\frac{f(\varepsilon, \ell_k)\,\Theta \left( 2\varepsilon - 2\varphi(r) - \frac{\ell^2}{r^2}\right)}{\hat{T} \left( \varepsilon, \ell_k \right)\sqrt{2\varepsilon - 2\varphi(r) - \frac{\ell^2}{r^2}}},
\end{equation}
where the following notation is introduced (we will denote different distribution functions by the same symbol if they can be distinguished by the number of arguments)
\begin{equation}\label{sp8}
f(\varepsilon, \ell_k)=\int d \tau \,d \gamma\, f(\varepsilon, \ell_k, \tau_l, \gamma)
\end{equation}
for distribution over only normalized energy and angular momentum.

Moving on, let us split the integration over normalized angular momentum $\ell_k$ in \eqref{rho4} on integration over its magnitude $\ell$ and integration over sphere $S_\ell$ with radius $\ell$:
\begin{equation}\label{rho5a}
\rho(r) = \frac{m}{2\pi r^2} \int d \varepsilon \, \int\limits_0^\infty d\ell\,
\frac{\hat f(\varepsilon, \ell)\,\Theta \left( 2\varepsilon - 2\varphi(r) - \frac{\ell^2}{r^2}\right)}{\hat{T} \left( \varepsilon, \ell \right)\sqrt{2\varepsilon - 2\varphi(r) - \frac{\ell^2}{r^2}}}.
\end{equation}
Here we replaced $\hat{T} \left( \varepsilon, \ell_k \right)$ with $\hat{T} \left( \varepsilon, \ell \right)$, because in spherically symmetric case the period doesn't depend on the direction of the angular momentum vector. We also introduced the following notation:
\begin{equation}\label{sp11}
\hat f(\varepsilon, \ell)=
\int \limits_{S_\ell} d^2 \ell  \,
f \left(  \varepsilon, \ell_k\right).
\end{equation}
The quantity $\hat f(\varepsilon, \ell)$ has the meaning of the particle distribution function over normalized energy and angular momentum modulus.

Let us consider the behavior of the obtained density \eqref{rho5a} in the limit $r \to 0$. Due to the fact that the gravitational potential is related to the non-negative density of matter by the equation \eqref{poisson}, the quantity $-\varphi(r)$ at $r \to 0$ cannot grow faster or in the same way as ${1}/{r^2}$. Hence, the term ${\ell^2}/{r^2}$ dominates in the argument of the $\Theta$-function at small $r$ and the only values which contribute to the integral over $\ell$ are the small values that satisfy the inequality
\begin{equation}
\ell\le r\sqrt{2\varepsilon - 2\varphi(r)}.
\end{equation}
As a result, the asymptotic behavior of the $\rho(r)$ at $r\to0$ is defined by the behavior of the functions $\hat f(\varepsilon, \ell)$ and $\hat{T}(\varepsilon,\ell)$, contributing to \eqref{rho5a}, at $\ell\to0$ .
The period of radial motion $\hat{T}(\varepsilon,\ell)$ has a finite limit $\hat{T}(\varepsilon,0)$ at $\ell\to0$ with a possible exception of small intervals of normalized energy $\varepsilon$ which correspond to the particles which are almost at rest in the center and whose contribution is negligible.
The behavior of $\hat f(\varepsilon, \ell)$ is not predetermined, so let us consider various options.

Assume, at first, that $\hat f(\varepsilon, \ell)$ has a finite limit at $\ell\to0$ and $\hat f(\varepsilon, 0)\ne0$ at least for some interval of $\varepsilon$.
In this case, we can perform a change of variables $\ell=r\tilde\ell$ in the integration in \eqref{rho5a}, which leads to the asymptotic behavior
\begin{equation}\label{rho5}
\rho(r) = \frac{m}{2\pi r} \int d \varepsilon \, d\tilde\ell\,
\frac{\hat f(\varepsilon, 0)\,\Theta \left( 2\varepsilon - 2\varphi(r) - \tilde\ell^2\right)}{\hat{T} \left( \varepsilon, 0 \right)
\sqrt{2\varepsilon - 2\varphi(r) - \tilde\ell^2}}=
\frac{m}{4 r} \int d \varepsilon \,\frac{\hat f(\varepsilon, 0)}{\hat{T} \left( \varepsilon, 0 \right)}.
\end{equation}
It can be seen that particles density is singular at $r\to0$ in this case, moreover, it is proportional to $1/r$ (note that $\hat f(\varepsilon, 0)\ge0$, so the integral in \eqref{rho5} does not vanish).
Such behavior exactly coincides with the one resulting from numerical simulations \emph{cusp}-like density profile with $\al=-1$ (see Introduction). It is worth noting that the gravitational potential $\varphi(r)$ in this case remains finite at $r=0$,
but has a cusp at this point, i.e. $\varphi'(0)\ne0$.

Now let us assume that distribution function $\hat f(\varepsilon, \ell)$ can be expanded into a series w.r.t. variable $\ell$ at the point $\ell=0$ and its zeroth term vanishes at any value of  $\varepsilon$, i.e. at $\ell\to0$
\begin{equation}\label{sp4}
\hat f(\varepsilon, \ell)\approx \hat f'(\varepsilon, 0)\ell,
\end{equation}
where prime means derivative w.r.t. $\ell$.
In this case, we get
\begin{equation}\label{rho6}
\rho(r) = \frac{m}{2\pi} \int d \varepsilon \, d\tilde\ell\,
\frac{\hat f'(\varepsilon, 0)\tilde \ell\,\Theta \left( 2\varepsilon - 2\varphi(r) - \tilde\ell^2\right)}{\hat{T} \left( \varepsilon, 0 \right)
\sqrt{2\varepsilon - 2\varphi(r) - \tilde\ell^2}}=
\frac{m}{2\pi} \int d \varepsilon \,\frac{\hat f'(\varepsilon, 0)}{\hat{T} \left( \varepsilon, 0 \right)}\sqrt{2\varepsilon - 2\varphi(r)}
\end{equation}
instead of \eqref{rho5} by analogous reasoning.
Assuming the finiteness of the value of the gravitational potential at zero $\varphi(0)$ (which is true even in the  \emph{cusp}-case), we can replace $\varphi(r)$ to $\varphi(0)$ in the last expression in \eqref{rho6}, when considering asymptotic behavior at $r\to0$.
It means that in this case the dependence of the density on radius corresponds to the \emph{core}-type profile.

To sum up, a rather simple analytical derivation shows that the matter density profile in the central region is defined by the presence or absence of a zeroth-order term in the expansion of the function $\hat f(\varepsilon, \ell)$ into series w.r.t $\ell$ at the point $\ell = 0$.
Since $\hat f(\varepsilon, \ell)$ is positive it is sufficient to check the expansion of the distribution function
\disn{sp15}{
\hat f(\ell)=\int d\varepsilon \hat f(\varepsilon, \ell)
\nom}
of particles  only over the module of angular momentum.
In the next section we will analyze the possible behavior of this function for $\ell\to0$.

\section{\boldmath An asymptotic behavior of the distribution function at $\ell\to0$}\label{razd2}
\subsection{The case of the spherically symmetric potential}\label{razd21}
First, let us consider the case when the formation of the static structure from a particles cloud goes in the already existing spherically symmetric gravitational potential $\ff(x_i)$. In this case, each particle preserves its angular momentum so the distribution function $\hat f(\ell)$ defined by \eqref{sp15} doesn't change with time.
Hence, it is sufficient to consider the distribution of particles at the initial moment of time in order to find the asymptotic of the distribution function in the resulting static configuration.

Let us start by obtaining the distribution function over the normalized energy and angular momentum $f(\varepsilon, \ell_k)$. At the first sight, this function should be smooth at $\ell_k=0$. Let us show that this is not the case in general.
Assume that we have a large number of point particles with coordinates $x_i$ and velocities $v_i$ which are described by the distribution function $\chi(x_i,v_i)$ at the initial moment. Then
\begin{equation}\label{sp10}
f(\varepsilon, \ell_k)=
\int d^3 x \, d^3 v\, \chi(x_i,v_i)
\delta(\ell_i-\epsilon_{ikl}x_k v_l)
\delta\left(\varepsilon-\frac{v^2}{2}-\ff(x_i)\right),
\end{equation}
where $v$ is the magnitude of velocity and $\epsilon_{ikl}$ is the Levi-Civita symbol.
We assume for simplicity that the function $\chi(x_i,y_i)$ is spherically symmetric which means that its value doesn't change under a simultaneous rotation of vectors $x_i$ and $v_i$.
Taking into account that $\ff(x_i)$ is also spherically symmetric, the function $f(\varepsilon, \ell_k)$ given by \eqref{sp10} can depend on $\ell_k$ only through its magnitude.
Then we can take $\ell_k = (\ell, 0, 0)$ without the loss of generality.
As a result, we have
\disn{sp9}{
f(\varepsilon, \ell_k)=\int d^3 x \, d^3 v \, \chi(x_i,v_i) \times\ns \times
\delta(\ell - x_2 v_3 + x_3 v_2)\delta(x_1 v_3 - x_3 v_1)\delta(x_1 v_2 - x_2 v_1) \delta\left(\varepsilon-\frac{v^2}{2}-\ff(x_i)\right)= \no
		= \int d x_2 \, d x_3 \, d^3 v \, \left.\left[\chi(x_i,v_i)\delta\left(\varepsilon-\frac{v^2}{2}-\ff(x_i)\right)\right]\right|_{x_1 = \frac{v_1}{v_3} x_3}\times\ns \times  \frac{1}{|v_3|}\, \delta(\ell - x_2 v_3 + x_3 v_2) \, \delta \left( \left( \frac{x_3 v_2}{v_3} - x_2 \right) v_1 \right) = \ns
		= \frac{1}{\ell}\int d x_2 \, d x_3 \, d v_2 \, d v_3 \, \left.\left[\chi(x_i,v_i)\delta\left(\varepsilon-\frac{v^2}{2}-\ff(x_i)\right)\right]\right|_{x_1 =v_1= 0}  \delta(\ell - x_2 v_3 + x_3 v_2).
\nom}
It is easy to check that after factorization of the multiplier $1/\ell$ the remaining expression becomes finite at $\ell=0$ in the general case.
This is true even if we lift the spherical symmetry restriction of the function $\chi(x_i,v_i)$. In this case
the limit of the coefficient before $1/\ell$ at $\ell\to0$ smoothly depends on the direction of the vector $\ell_k$.

As a result, we see that $f(\varepsilon, \ell_k)$ is indeed non-smooth at $\ell_k=0$.
In this case, the distribution function \eqref{sp15} over the module of angular momentum behaves at $\ell\to0$ as
\disn{sp11a}{
\hat f(\ell)\approx C\ell,
\nom}
where $C$ is obtained by integrating over $\varepsilon$ and averaging over the angles of the abovementioned limit of the coefficient before $1/\ell$ in \eqref{sp9}.
Since it corresponds to \eqref{sp4}, we can conclude that the \emph{core}-type matter density profile emerges in the considered case.

The discussed situation when the gravitational potential remains spherically symmetric during the process of the density profile formation may take place if the gravitational potential is created primarily by the dark matter.
At the same time, dark matter can obey its own laws (for example, it may have some specific self-interaction) which define whether it forms \emph{cusp} or \emph{core}. Regardless of this, here we studied the formation of a density profile for a regular matter and obtained the \emph{core}-type result.

\subsection{The case with a deviation from the spherical symmetry}\label{razd22}
Now let us switch to the alternative situation when the static structure is formed by a cloud of particles without a spherically symmetric gravitational background. Let us assume that gravitational potential forms simultaneously with the static configuration, so a significant deviation from spherical symmetry may occur during that process.
This situation might take place if we consider the formation of structures from dark matter particles assuming that it behaves as regular matter without any self-interaction. Such a setup is close to the one used in mentioned numerical simulations (see Introduction).

In this case trajectories of particles would not be exactly as shown in the figure~\ref{pic1} and the particle's position would not be defined by parameters $\varepsilon, \ell_k, \tau_l, \gamma$. However, if the deviation from spherical symmetry is small then deviations from defined trajectories will be also small and particles movement can still be described by parameters $\varepsilon, \ell_k, \tau_l, \gamma$, but we should take into account that not all of these parameters are conserved with time.
In particular, the angular momentum of each particle, determined w.r.t. to the future center of the emerging structure, will no longer conserve its original value, since it can change under the action of forces arising from small conglomerates of particles that can evolve into satellite galaxies.
As a result, in contrast with the previous case, the distribution function $f(\varepsilon, \ell_k, \tau_l, \gamma)$, and hence $f(\varepsilon,\ell_i)$, can change over time, including the change in terms of distribution over $\ell_i$.
It means that the distribution function over the module of angular momentum defined by formulas \eqref{sp11},\eqref{sp15} can also change and its asymptotic behavior at $\ell=0$ determines whether the \emph{core} profile or the \emph{cusp} profile arise.

As shown in the previous subsection, this function has asymptotic behavior \eqref{sp11a} at the initial moment in the general case. However, in contrast with the previous case, the function $\hat f(\ell)$ can now change with time, as well as its asymptotics. Therefore this asymptotics can acquire a non-zero contribution in a form of a zeroth-order term in its expansion into series over $\ell$, which will change the emerging density profile type to \emph{cusp}.

The rate of change of the normalized angular momentum $\ell_i$ for a single particle is determined by the moment of force acting on the particle:
\disn{sp12}{
\dot \ell_i = \epsilon_{ikl} x_k a_l,
\nom}
where $x_i$ is the location of the particle at a given moment of time, and $a_l$ is its acceleration; dot denotes derivative w.r.t. time.
Hence, for the modulus of the normalized angular momentum $\ell$ of a single particle, we obtain the expression
\disn{sp13}{
\dot \ell = \frac{d}{dt} \sqrt{\ell_i \ell_i} = \varepsilon_{ikl}\frac{\ell_i}{\ell}x_k a_l,
\nom}
which can be either positive or negative.

Time evolution of the distribution function $\hat f(\ell)$ can be described by a standard continuity equation
\disn{sp14}{
\frac{d}{dt}\hat f(\ell)=-\frac{d}{d\ell}\ls \hat f(\ell)\bar{\dot\ell}\,\rs,
\nom}
where $\bar{\dot\ell}$ is a value of the rate of change $\dot\ell$ averaged over all particles with given $\ell$. So, the $\hat f(\ell)\bar{\dot\ell}$ is a "flow"{} of particles in the space of values of the angular momentum modulus. This "flow"{} can be either positive or negative. A positive value means "outflow"{} of particles from the given point $\ell$ of modulus of the specific angular momentum. A negative value means "inflow"{} to this point.
A negative value of this ''flow''{} at $\ell = 0$  after some time should lead to the appearance of a non-zero value $\hat f(0)$.
It can be seen more precisely from equation \eqref{sp14} if we substitute \eqref{sp11a} into it (note that $C>0$ because the distribution function $\hat f(\ell)$ is positive) as an initial value and neglect the change of $\bar{\dot\ell}$ with changing $\ell$ near the point $\ell=0$.
If $\bar{\dot\ell}<0$ we obtain the solution, which at $\ell>0$ has the form
\disn{sp15a}{
\hat f(\ell)=C(\ell-\bar{\dot\ell}\,t),
\nom}
and also contain a delta-functional term at $\ell=0$. In reality, such contribution doesn't arise if we consider the dependence of the value $\bar{\dot\ell}$ on $\ell$ at $\ell\to0$.
It is expected that unless a fine-tuning was made the value $\hat f(\ell)$ in that asymptotic becomes nonzero after some time $t$.
Moreover, it can tend to a finite value $\hat f(0)>0$ or, probably, tend to infinity as well.

As it can be seen from \eqref{sp13}, particles with opposite values of a normalized angular momentum $\ell_i$ give the opposite contributions to the $\bar{\dot\ell}$ at $\ell\to0$. If there were exactly the same number of particles of both types, then this would give a zero value of $\bar{\dot\ell}$ (note that, as can be seen from \eqref{sp9}, the distribution of particles over the vector $\ell_i$ is not smooth at $\ell\to0$, which reduces the reliability of this conclusion).
However, if the spherical symmetry is violated and the system has a certain angular momentum then the number of such particles wouldn't be the same, which may lead to the appearance of a non-zero $\bar{\dot\ell}$.

It is too difficult to determine exactly which sign will acquire the average value over all $\bar{\dot\ell}$ particles as a result of spherical symmetry violation.
However, it is natural to assume that in a generic situation there are intervals of time when $\bar{\dot\ell}<0$. During such a period of time, the quantity $\hat f(0)$ has a non-zero value, which may change with further dynamics, but without fine-tuning of the initial data it will not disappear completely. Thus, we have significant arguments supporting the fact that in the presence of deviations from spherical symmetry in general case after some time the distribution function
$\hat f(\ell)$ over the modulus of a normalized angular momentum at $\ell\to0$ will no longer behave according to \eqref{sp11a}.

If we assume that $\hat f(\ell)$ can be expanded into a series w.r.t. variable $\ell$ at the point $\ell=0$,
then
\disn{sp17}{
\hat f(\ell)\approx \hat f(0)+\hat f'(0)\ell
\nom}
with $\hat f(0)>0$. As shown in section~\ref{razd1}, this corresponds to the matter profile of the \emph{cusp} type
with $\al=-1$.
However, the function $\hat f(\ell)$ doesn't have to be analytical at zero. Instead, it can (as was mentioned before) tend to infinity at $\ell\to0$ with a power law behavior with a non-integer exponent.
The restrictions on the value of this exponent will be discussed in the next section.

\section{Deviations from the complete spherical symmetry}\label{razddop}
Before in section \ref{razd1}, we assumed that matter distribution and hence a gravitational potential have spherical symmetry.
For real galaxies is usually not the case, so it is important to examine how the results change if we assume that spherical symmetry no longer holds exactly.
As it was mentioned in the section \ref{razd22}, while the deviations from exact symmetry are small, trajectories of particles can still be described by parameters $\varepsilon, \ell_k, \tau_l, \gamma$, but the normalized angular momentum $\ell_k$ will no longer conserve.

Without spherical symmetry the formula \eqref{rho2} (and also \eqref{rho3}) is still correct, if we consider that $\rho(r)$ is a matter density averaged over angles. We are interested in this quantity when we study a radial distribution of matter without spherical symmetry.
The estimate \eqref{sp3} and hence the formula \eqref{rho4} remain correct after the restriction of the time interval $T$ so that $T\gg \hat{T} \left( \varepsilon, \ell \right)$ but the change of normalized angular momentum $\ell$ is sufficiently small. The consistency of these requirements imposes restrictions on the amount of deviation from spherical symmetry when the formula \eqref{rho4} remains sufficiently accurate.
Performing in \eqref{rho4} the change of variables $\ell_k=r\tilde\ell_k$ analogous to the one used in the obtaining of \eqref{rho5}, we have
\disn{spn1}{
\rho(r) = \frac{mr}{2\pi} \int d \varepsilon \, d^3\tilde\ell\,
\frac{f(\varepsilon, r\tilde\ell_k)\,\Theta \left( 2\varepsilon - 2\varphi(r) - {\tilde\ell^2}\right)}{\hat{T} \left( \varepsilon, r\tilde\ell_k \right)\sqrt{2\varepsilon - 2\varphi(r) - {\tilde\ell^2}}}.
\nom}
Again, at $r\to0$ we obtain the result that the asymptotic of $\rho(r)$ is determined by the behavior of the distribution
function $f(\varepsilon, \ell_k)$ at small values of $l_k$ even in the absence of spherical symmetry. Moreover
\disn{spn1.1}{
\rho(r)\sim r f(\varepsilon, r\tilde\ell_k).
\nom}

In section~\ref{razd21} a setup was considered where the static structure is forming in a given gravitational potential.
Now we assume that, in contrast with the previous case, the gravitational potential isn't exactly spherically symmetric, but still has axial symmetry.
Such a symmetry provides a better fit to the properties of real galaxies.
In the presence of only axial symmetry, the only conserving component of angular momentum is $\ell_3$ if the coordinate $x^3$ is aligned with the symmetry axis.
As a consequence, the distribution function $f(\varepsilon, \ell_k)$ changes over time but only in the part related to distribution over $\ell_{||}=\sqrt{l_1^2+l_2^2}$. As a result, a ''flow''{} of particles in the space of values of $\ell_{||}$ analogous to the one mentioned in~\ref{razd22} can arise, but now not all components of angular momentum can vary. At the same time the value
\disn{spn2}{
\int d^2 \ell_{||}\, f(\varepsilon, \ell_k),
\nom}
i.e. the distribution over the component of the normalized angular momentum $\ell_3$ should remain unchanged in time.

At the initial moment of time we have $f(\varepsilon, \ell_k)\sim 1/\sqrt{\ell_3^2+\ell_{||}^2}$ at $l_k\to0$ corresponding to \eqref{sp9} even in the presence of deviations from spherical symmetry (see the text after \eqref{sp9}).
The fastest growth of the distribution function $f(\varepsilon,\ell_k)$ at each value of $\ell_3$ and $\ell_{||}\to0$ appearing as a result of ''inflow''{} of particles into the point $\ell_{||}=0$ (again in the sense discussed in section~\ref{razd22}) should correspond to the finiteness of the integral \eqref{spn2}.
It means that the fastest growth can be estimated by a power law as $f(\varepsilon, \ell_3,\ell_{||})\sim 1/\ell_{||}^\be$, where $\be<2$.
Using this estimate in \eqref{spn1.1} we obtain that the fastest growth of density at $r\to0$ can have the form $\rho(r)\sim 1/r^{\be-1}$. Since the density of matter has a finite central value at the initial moment, we can conclude that $\rho(r)\sim r^{\al}$ with $\al>-1$ if the formation of static structure goes in the gravitational potential which has axial symmetry.
It corresponds to the \emph{cusp} profile instead of \emph{core}, but with a less steep growth than $1/r$.

Now consider the case when the gravitational potential does not even have axial symmetry.
Then there are no conservation of the angular momentum $\ell_k$
(and hence the results will not change if we switch to the case discussed in section~\ref{razd22} where the gravitational potential forms simultaneously with the density profile).
As a result, the  dependence on all components of $\ell_k$ of the distribution function $f(\varepsilon, \ell_k)$ can change with time, and hence the ''flow''{} may arise in the space of all components of the angular momentum.
Then the fastest growth of the distribution function $f(\varepsilon,\ell_k)$ at $\ell_k\to0$, which arises as a result of possible ''inflow''{} of particles into the point $\ell_k=0$, is restricted only by the condition of integrability of distribution function over all components $\ell_k$.
Furthermore, the fastest growth can be estimated by a power law as $f(\varepsilon, \ell_k)\sim 1/\ell^\be$, where $\be<3$.
Using this estimate in \eqref{spn1.1}, we obtain that the growth of the density at $r\to0$ should have the form $\rho(r)\sim r^{\al}$ with $\al>-2$ even in the absence of axial symmetry.
This statement holds both in the case of the formation of the static structure in existing asymmetric gravitational potential and in the case when the gravitational potential is formed simultaneously with the density profile.

\section{Conclusion}
We have analyzed the formation of a static density profile $\rho(r)$ of a radial matter distribution arising for a non-interacting (with an exception of gravitation interaction) dust-like matter.
We established a connection \eqref{spn1.1} of the type of arising profile with the asymptotic behavior of the distribution function $f(\varepsilon, \ell_k)$ of particles over normalized energy and angular momentum  at $\ell_k=0$.
Particularly, in terms of the distribution function $\hat f(\ell)$ over the module of the normalized angular momentum the relation takes the following form: if the function $\hat f(\ell)$ tends to a finite nonzero value at $\ell\to0$ then the \emph{cusp} type profile arise with $\rho\sim r^{\al}$, $\al=-1$. In another case when $\hat f(\ell)\sim \ell$, the \emph{core} type profile arise instead.

The determination of the asymptotic behavior of distribution functions for $\ell_k=0$ is carried out for situations where the resulting stationary configuration has either exact spherical symmetry or some deviations from it. The latter case can be more suitable in the modeling of galaxy formation.

Two possible scenarios for the formation of a density profile can be considered.
In the first scenario, we assume that first a potential well with a certain symmetry (the gravitational potential $\ff(r)$) somehow arises.
For example, it can be formed by dark matter, the properties of which may differ from the properties of ordinary dusty matter due to the presence of some self-interaction. We study the formation of the distribution of ordinary matter with a given background potential. Speaking of the symmetry of the potential well, we first of all study the case of exact spherical symmetry, which provides the most rigorous results.
In this case $\hat f(\ell)\sim \ell$ and \emph{core} type profile arise.
If at the same time the self-interaction of dark matter also provides a \emph{core} distribution for it then this case is consistent with the majority of observations.
Further, we discuss how the results change with the presence of deviations from spherical symmetry.

If the deviations are that the exact axial symmetry present then it is possible to provide arguments supporting the origin of the values $\al>-1$, i.e. the \emph{cusp} profile with density growth slower than $1/r$ can arise as well as the \emph{core} profile.
And if the axial symmetry is also broken then values $\al>-2$ are also possible. It means that the \emph{cusp} profile with density growing as $1/r$ (this case is distinguished because it corresponds to the analytical behavior \eqref{sp17} of the distribution function of particles over the module of angular momentum) or even faster can arise.

The same result occurs in the second scenario of the density profile formation. Here we assume that there is no pre-formed potential well, and the gravitational potential is formed simultaneously with the distribution profile of dust-like matter. Since in this case noticeable deviations from spherical symmetry occur in the formation of a stationary configuration, the result is the same as with complete symmetry breaking in the first scenario. The second scenario is suitable for describing the formation of the density profile of both dark and ordinary matter. The result obtained in this case is consistent with the results of numerical simulations giving the \emph{cusp}-profile, in which the problem statement is close to the described second scenario.

{\bf Acknowledgments.}
The authors are grateful to A.~Golovnev for useful discussions
and A.~El~Zant for provided references.
The work is supported by RFBR Grant No.~20-01-00081.
The work of A.D.~Kapustin is supported by the Foundation for the Advancement of Theoretical Physics and Mathematics "BASIS"{}.


\end{document}